\documentclass{article}
\usepackage{jsaisig}
\usepackage{amsfonts}
\usepackage{amssymb}
\usepackage{amsmath}
\usepackage{amsthm}
\usepackage{mathtools}
\usepackage{multirow}
\usepackage{booktabs} 
\usepackage{url}

\title{A Constraint Programming Approach to Fair High School Course Scheduling} 

\author{
Mitsuka Kiyohara\afil{1}
\quad 
Masakazu Ishihata\afil{2} 
}

\affiliation{%
\afil{1} University of Toronto
\quad 
\afil{2} NTT Communication Science Laboratories
}

\abstract{
    Issues of inequity in U.S. high schools' course scheduling did not previously exist. 
    However, in recent years, with the increase in student population and course variety, students perceive that the course scheduling method is unfair. 
    Current integer programming (IP) methods for the high school scheduling problem (HSSP) fall short of addressing these fairness concerns. 
    The purpose of this research is to develop a solution methodology that generates feasible and fair course schedules using student preferences. 
    Utilizing principles of fairness, which have been well studied in market design, we define the fair high school scheduling problem (FHSSP), a novel extension to the HSSP, and devise a corresponding algorithm based on integer programming to solve the FHSSP.
    We test our approach on a real course request dataset from a high school in California, USA.
    Results show that our algorithm can generate schedules that are both feasible and fair. 
    In this paper, we demonstrate that our IP algorithm not only solves the HSSP and FHSSP in the United States but has the potential to be applied to various real-world scheduling problems.
    Additionally, we show the feasibility of integrating human emotions into mathematical modeling. 
}

\newtheorem{example}{Example}

\newtheorem{definition}{Definition}

\begin{document}
\maketitle
\pagestyle{plain}

\section{Introduction}
In today’s landscape of high school education in the United States, schools face challenges in scheduling due to the increase in student population and the wide variety of course offerings. 
More than a decade ago, there were a total of 16,040 public and private high schools with 16 million students. 
Today, this number has increased with the addition of 912 schools and 1 million students \cite{nces_2022}. As a result, ensuring efficient scheduling in high schools has become increasingly difficult to maintain. 
At the same time, the recognition of existing inequity in course schedules among students has come forward.

The specifications of high school scheduling vary significantly across different countries due to diverse cultural and educational settings, making it difficult to develop a generalized solution. 
For instance, the United States adopts a decentralized education system, where each state has its own set of educational standards and curriculum guidelines. 
This decentralization makes the problem complex, as it leads to a wide range of course offerings, course requests, and constraints from both students and the school. 

Course schedules for high schools are commonly generated using a constraint programming (CP) solver. 
However, solvers in current use do not consider fairness as a constraint; therefore, this method is efficient but lack fairness. 
Another approach to solve the HSSP is to use an existing solver and then manually modify the schedule to ensure fairness. However, this method is inefficient and is prone to human error. 
Hence, we conclude that an efficient and fair CP solver is needed. 

A rich literature exists in the field of market design regarding matching with fairness constraints, such as the daycare matching problem, course allocation problem, and school choice problem. 
Yet, very little research has been done on the HSSP, which may be due to the complexity of the problem and the large number of constraints.
Since fairness is a well-studied principle in market design, we decide to incorporate these concepts as a constraint into an integer programming (IP) model. 
We are the first to study the HSSP in which fairness is defined (FHSSP), as well as develop and implement an algorithm to solve the problem described above. 

The rest of the paper is organized as follows. 
First, we review related research (Section 2). 
Then, we formally state the HSSP in the United States (Section 3) and the new FHSSP model (Section 4). 
Later, we detail the IP model (Section 5) and explain the experimental setup and results (Section 6). 
Finally, we conclude the paper and discuss future work (Section 7).  

\section{Related Work}
The formulation of the high school scheduling problem using IP has been explored in the literature, notably by \cite{kristiansen_integer_2015} and \cite{ribic_modelling_2015}. These studies, however, do not include fairness constraints, a crucial aspect of our research. In contrast, the use of IP methods for course scheduling in universities, along with real-life data implementation, has been addressed by \cite{durakbasa_course_2023} and \cite{tran_integer_2021}, although their focus differs from our high school-centered approach.

In the field of market design, fairness, particularly stability in allocation, has been extensively studied with a focus on various matching markets. Algorithms aiming to compute stable matchings based on agents' preferences have been developed, as seen in the work of \cite{sun2022daycare}, \cite{somnez2003}, and \cite{hamada2011}. However, these have not been directly applied to high school course scheduling scenarios.
The student course allocation market, a problem closely related to course scheduling, has been examined in university settings by \cite{budish_multi-unit_2012}, \cite{diebold_course_2014}, and \cite{kotsireas_student_2019}. These studies, while insightful, focus on different constraints and do not specifically address high school environments. Our research bridges this gap by extending the concept of fairness to high school course scheduling and adapting it to unique constraints in US high schools.

In summary, while existing literature provides a solid foundation in both course scheduling and fairness, our research stands out in its application of these concepts to the high school setting, with an emphasis on fairness and a novel set of constraints. 

\section{High School Scheduling Problem (HSSP)}
\label{sec:HSSP}

In this section, we formulate the \emph{high school scheduling problem} (HSSP) as a constraint satisfaction problem (CSP) to assign students appropriately to each course. 
We first introduce the input and decision variables of this problem and then define a feasible schedule using the introduced variables.

\subsection{Notation}
\label{ssec:notation}

There are a set of courses $C$, a set of instructors $I$, and a set of students $S$. 
A \emph{lecture} is an ordered pair of a course $c \in C$ and instructor $i \in I$. 
Note that students are assigned to lectures, not directly to courses. 
For instance, when a student $s$ is enrolled in lecture $l = (c, i)$, we say $s$ takes course $c$ taught by instructor $i$.

Let $W$, $P$, and $R$ be a set of weekdays (Monday, \dots, Friday), school periods per day ($1,\dots,|P|$), and rooms, respectively.
A \emph{time slot} is an ordered pair consisting of a weekday $d \in W$ and school period $p \in P$. 
A \emph{unit} is a tuple consisting of a time slot $t \in T \triangleq W \times P$, room $r \in R$, and lecture $l$, where $l$ is scheduled in $r$ at $t$.

Each instructor $i$, student $s$, and room $r$ have associated sets of \emph{eligible courses}, denoted as $E_i$, $E_s$, and $E_r$, respectively. 
$E_i$ represents the set of courses that instructor $i$ is certified to teach, $E_s$ represents the set of courses that student $s$ is qualified to take, and $E_r$ represents the set of courses that can be hosted in room $r$, considering the room's facilities.

For any $x \in \{i, s, r\}$, $N_x^-$ (resp. $N_x^+$) is the minimum (resp. maximum) number of units $i$ is required to teach per week, the number of courses student $s$ can enroll this semester, and the number of students who can attend room $r$ at any given unit, respectively.
We define \emph{lecture frequency} for a given lecture $l = (c, i)$ as the total number of units $u = (c, i, r, t)$ scheduled for $l$, and $N_c$ indicates the lecture frequency for a course $c$.
All the aforementioned variables are inputs to the HSSP.

We then introduce three decision variables $L$, $A$, and $U$ to construct a solution to this problem (i.e., a weekly school schedule). 
$L \subseteq C \times I$ describes the set of lectures, where $(c, i) \in L$ indicates that instructor $i$ teaches course $c$.
$A \subseteq L \times S$ represents the lecture-student assignments, where $(c, i, s) \in A$ indicates that student $s$ is enrolled in lecture $(c, i)$.
$U \subseteq T \times R \times L$ defines the set of units, where unit $(c, i, r, t) \in U$ indicates that lecture $(c, i)$ occurs in room $r$ during time slot $t$.
Thus, a weekly school schedule consists of a triple $\mathcal{S} \triangleq \langle A, L, U\rangle$.

\subsection{Feasible Schedules} 
\label{ssec:feasibility}

To define the feasibility of schedules, we establish three constraints.
The goal of the HSSP is to find \emph{feasible} schedule $\mathcal{S}$ satisfying the following constraints.

\paragraph{Time slot conflicts of instructors, students, and rooms} 

Any instructor $i$, student $s$, and room $r$ cannot be assigned two or more units in the same time slot.
We introduce $U_s$, $U_i$, and $U_r$ as follows:
\begin{enumerate}
\item 
$U_s \triangleq \{u \in U \mid \exists l \in L, (l,s) \in A \wedge l \in u \}$:
all units student $s$ partakes in
\item 
$U_i \triangleq \{u \in U \mid \exists l \in L, i \in l \wedge  l \in u\}$: 
all units instructor $i$ teaches
\item 
$U_r \triangleq \{u \in U \mid r \in u \}$: 
all units offered in room $r$
\end{enumerate}
Then, $U_x$ ($x \in \{s, i, r\}$), which is the set of units assigned to $x$,  must satisfy the following equation for any time slot $t$:
\begin{align}
\forall \{u, u'\} \subseteq U_x, \neg (t \in u \wedge t \in u')
\label{eq:cons:tsc}
\end{align}

\paragraph{Eligibility of instructors, students, and rooms} 

The appropriate instructors, students, and rooms must be assigned for all courses $C$.
For any instructor $i$, student $s$, and room $r$, $C_x$ $(x \in \{s, i, r\}$) is defined as follows:
\begin{enumerate}
\item 
$C_s \triangleq \{c \in C \mid \exists l \in L, (l,s) \in A \wedge c \in l\}$:
all courses student $s$ takes
\item 
$C_i \triangleq \{c \in C \mid (c, i) \in L\}$:
all courses instructor $i$ teaches
\item 
$C_r \triangleq \{c \in C \mid \exists u \in U_r, \exists l \in L, l \in u \wedge c \in l\}$:
all courses offered in room $r$
\end{enumerate}
Then, $C_x$, which is the set of courses assigned to $x$, must satisfy the following constraint:
\begin{align}
C_x &\subseteq E_x
\label{cnst:eligibility}
\end{align}
%

\paragraph{Numerical Constraints} 

Each instructor $i$, student $s$, and room $r$ is given the appropriate range of the number of units $i$ teaches per week, the number of courses $s$ takes this semester, and the number of students who may attend a unit at $r$, respectively.
Let $U_{i,d} \triangleq \{ u \in U_i \mid \exists t \in T, d \in t \}$ be the set of units on $d$ that instructor $i$ teaches and $S_u \triangleq \{s \in S \mid u \in U_s\}$ as the set of all students who partake in unit $u$. Then, $U_{i,d}$,  $C_s$, and $S_u$ must satisfy the following constraints, where $r \in u$: 
\begin{align}
    N_i^- &\leq |U_i| \leq N_i^+
    \label{cnst: inst_limit}
    \\
    N_s^- &\leq |C_s| \leq N_s^+  
    \label{cnst: course_limit}
    \\
    N_r^- &\leq |S_u| \leq N_r^+
    \label{cnst: room_cap}
\end{align}
%
\paragraph{Lecture frequency} 

Any lecture $l$ must be offered for a predefined frequency per week.
Let $U_l \triangleq \{u \in U \mid l \in u \}$ be the number of units corresponding to $l$.
Then, $U_l$ must satisfy the following constraint, where $c \in l$:
\begin{align}
|U_l| = N_c
\label{cnst:lec_frequency}
\end{align}

Therefore, for a schedule to be \emph{feasible}, all of the previous constraints (\eqref{eq:cons:tsc}, \eqref{cnst:eligibility}, \eqref{cnst: inst_limit}, \eqref{cnst: course_limit}, \eqref{cnst: room_cap}, and \eqref{cnst:lec_frequency}) must hold. 

\section{Fair High School Scheduling Problem (FHSSP)}
\label{sec:FHSSP}

In this section, we formulate the \emph{fair high school scheduling problem} (FHSSP), an extension to the HSSP which contains fairness constraints.
We define two key quantities, the \emph{degree of interest} and \emph{priority} for each student-course pair, and subsequently formulate the \emph{fairness} constraint using the introduced quantities.

\paragraph{Degree of interest}
Let $d_{s,c}$ be the \emph{degree of interest} of student $s$ for course $c$. 
Here, we assume $d_{s,c}$ is \emph{normalized}. 
Namely, for any student $s \in S$: 
\begin{equation}
    0 \leq d_{s,c} \leq 1 \wedge \sum_{c \in C} d_{s,c} = 1
\end{equation}
In other words, $\{d_{s,c} \}$ represents the student preference. 

\paragraph{Priority order}
For any student $s$ and course $c$, let $p_{s,c}$ be the \emph{priority} for $s$ to take $c$, which is a pre-determined value based on student $s$'s past academic achievement and preference. 
If $c$ has one or more prerequisites, $p_{s,c}$ is determined by $s$'s average grade for those prerequisites. 
The higher the grade $s$ has, the higher priority $s$ receives for $c$. 
If $c$ does not require a prerequisite, then we resort to student $s$'s degree of interest. 
The higher the degree of interest $s$ has for $c$, the higher priority $s$ receives for $c$. 
Note that a student's priority order depends on the student's eligibility for a course (i.e., $e_{s,c} \implies p_{s,c}$).  
We calculate $p_{s,c}$ as follows: 
\begin{equation}
p_{s, c} =
    \begin{cases}
        \frac{1}{|C_p|}\sum_{c^\prime \in C_p} g_{s, c^\prime} & \text{if $c$ has prerequisite(s) $C_p$} \\
        d_{s,c} & \text{if $c$ has no prerequisite} \\ 
    \end{cases}
\end{equation}
where $C_p$ represents the set of prerequisite courses for $c$ and $g_{s, c'}$ as the grade of $s$ for $c'$. 

Introducing $p_{s,c}$ and $d_{s,c}$ allows students to compare their eligibility for a course with other students. 
We will illustrate this with an example: 
\begin{example}
    Consider two students $s$ and $s'$ with $p_{s,c}$, $d_{s,c}$ and $p_{s', c}$, $d_{s',c}$ respectively. 
    If $p_{s, c} > p_{s', c}$, then $s$ has a higher priority to take $c$ than $s'$. 
    Similarly, if $d_{s,c} > d_{s',c}$, then $s$ is more interested in taking $c$ than $s'$. 
    If both cases are true, then we conclude that student $s$ is more eligible to take course $c$ than student $s'$. 
\end{example}
We require both instances to be true because this condition produces a clear answer to why a student deserves a seat over another student.
Let us provide a counterexample: 
\begin{example}
    Assume $c$ is a course with prerequisites. 
    Suppose $p_{s, c} < p_{s', c}$ but $d_{s,c} > d_{s', c}$. 
    Then, the question of who deserves the seat more between $s$ and $s'$ becomes subjective. 
    One could argue that the student more academically prepared to take $c$ should be assigned to $c$, while another could argue that the student who shows more interest should be able to take $c$. 
    Therefore, we impose both quantities for one student to be either greater or smaller than the other student. 
\end{example}
\subsection{Fair Schedules}
The challenge of this problem is the potential for \emph{student envy}, which can be formally defined by a list of constraints. 
We propose this concept in the definition below.
\begin{definition}
    Consider two students $s$ and $s'$ and their respective course assignments $C_s$ and $C_{s'}$. Student $s$ is said to experience envy towards $s'$ if the following four conditions are true: \\ 
    (i) $c \in C_{s'} \setminus C_s$, (ii) $p_{s,c} > p_{s',c}$, (iii) $d_{s,c} > d_{s',c}$, and (iv) $|C_s| < |C_{s'}|$. 
    \label{cnst:fairness}
\end{definition}
Definition \ref{cnst:fairness} describes two students $s$ and $s'$ and a course $c$. Condition (i) states that $c$ is assigned to $s'$ but not to $s$. 
Condition (ii) states that $s$'s priority for $c$ is higher than that of $s'$. 
Similarly, condition (iii) states that $s$'s degree of interest for $c$ is greater than that of $s'$. 
Lastly, condition (iv) states that the number of courses assigned to $s$ is fewer than those assigned to $s'$. 

Recall that for the HSSP to return a feasible outcome, Constraints \eqref{eq:cons:tsc} to \eqref{cnst:lec_frequency} must hold. 
If the HSSP returns a feasible outcome, the FHSSP evaluates whether an optimal outcome that upholds fairness under the same problem setting exists. 
In other words, in addition to Constraints \eqref{eq:cons:tsc} to \eqref{cnst:lec_frequency}, the following fairness constraints must also hold: 
\begin{multline}
    (\exists c \in C_{s'} \setminus C_s) \wedge (p_{s,c} > p_{s',c}) \\ \wedge (d_{s,c} > d_{s',c}) \wedge (|C_s| < |C_{s'}|)
\end{multline}

\section{Proposed method}
In this section, we present an algorithm based on integer programming (IP) to solve the HSSP and FHSSP. 

\subsection{Binary variables}

First, we introduce three types of variables that are encoded into binary variables, starting with the decision variables ($L$, $A$, $U$). 
For every course $c \in C$ and instructor $i \in I$, create a variable $l_{c,i}$ indicating whether instructor $i$ is teaching course $c$ as follows: 
\begin{equation}
l_{c, i}=
    \begin{cases}
        1 & i \text{ is teaching } c\\
        0 & \text{otherwise } 
    \end{cases}
\end{equation}
For every student $s \in S$ and lecture $l = (c,i) \in L$, create a variable $a_{c, i, s}$ indicating whether student $s$ is enrolled to lecture $(c, i)$. 
\begin{equation}
a_{c, i, s}=
    \begin{cases}
        1 & s \text{ is enrolled in } (c,i)\\
        0 & \text{otherwise } 
    \end{cases}
\end{equation}
For every lecture $(c, i) \in L$, room $r \in R$, and time slot $t \in T$, create a variable $u_{c, i, r, t}$ indicating whether lecture $l$ is being offered in room $r$ on time slot $t$. 
\begin{equation}
u_{c, i, r, t}=
    \begin{cases}
        1 & (c, i) \text{ is offered in } r \text{ at time slot } t\\
        0 & \text{otherwise } 
    \end{cases}
\end{equation}

We also introduce auxiliary binary variables, which are later used to encode the constraints in this problem. 
For each student $s \in S$, instructor $i \in I$, and room $r \in R$, create a variable $c_{x,u}$ (where $x \in \{s, i, r\}$) indicating the participation of $x$ in course $c$. The variable is defined as:  
\begin{equation}
c_{x, c}=
    \begin{cases}
        1 & x \text{ is involved in } c\\
        0 & \text{otherwise } 
    \end{cases}
\end{equation}
%
For each student $s \in S$, instructor $i \in I$, and room $r \in R$, create a variable $u_{x,u}$ (where $x \in \{s, i, r\}$) indicating the assignment of $x$ in unit $u$. The variable is defined as:  
\begin{equation}
u_{x, u} =
    \begin{cases}
        1 & x \text{ partakes in } u\\
        0 & \text{otherwise } 
    \end{cases}
\end{equation}
%
For each student $s \in S$, instructor $i \in I$, and room $r \in R$, create a variable $e_{x,c}$ (where $x \in \{s, i, r\}$) indicating the eligibility related to course $c \in C$: 
\begin{equation}
e_{x, c} =
    \begin{cases}
        1 & x \text{ is eligible to take } c\\
        0 & \text{otherwise } 
    \end{cases}
\end{equation}
For every student $s \in S$, create a variable $\alpha_{s, c}$ indicating that student $s$ feels envy towards another student $s'$. 
\begin{equation}
\alpha_{s, s'} =
    \begin{cases}
        1 & s \text{ experiences envy towards } s'\\
        0 & \text{otherwise } 
    \end{cases}
\end{equation}

\subsection{Constraints}
Given the binary variables encoded in the previous section, we first show the encoding of the original problem constraints.  
For Constraint \eqref{eq:cons:tsc}, we can capture the requirement by the following: 
\begin{align}
    \sum_{u \in U_x} \sum_{t \in u} u_{x,u} \leq 1 \text{ } \forall x \in \{s, i, r\}  
\end{align}
Constraint \eqref{cnst:eligibility} corresponds to the set of eligible courses for each respective student, instructor, and room. The following must hold for all: 
\begin{align}
  \sum_{c \in C_x} c_{x, c} \leq \sum_{e \in E_x} e_{x,c} \text{ } \forall x \in \{s, i, r\}
\end{align}
Recall Constraint \eqref{cnst: inst_limit}, \eqref{cnst: course_limit} and \eqref{cnst: room_cap}, which represent the range on the number of units an instructor is allowed to teach per day, the number of courses a student can enroll in, and the number of students who may attend a unit for every room, respectively. 
For all $u \in U$ and $c \in C$, 
\begin{align}
    N_i^- \leq \sum_{i \in I} u_{i, u} \leq N_i^+ \\ 
    N_s^- \leq \sum_{c \in C} c_{s, c} \leq N_s^+ \\ 
    N_r^- \leq \sum_{s \in S} u_{s, u} \leq N_r^+  
\end{align}

Constraint \eqref{cnst:lec_frequency} represents the predefined frequency $N_c$ lecture $(c, i)$ must meet per week. Then, for all $(c, i) \in L$, 
\begin{align}
    \sum_{\forall r \in R} \sum_{\forall t \in T} u_{c,i,r,t} = N_c
\end{align}

Recall that a student $s$ experiences envy towards another student $s'$ if (i) $c$ is assigned to $s'$ but not to $s$; (ii) the priority for $s$ to take $c$ is higher than $s'$; (iii) $s$'s degree of interest to take $c$ is greater than $s'$; and (iv) the number of courses assigned to $s$ is fewer than $s'$ (see Definition \ref{cnst:fairness}). 
For all students $s, s' \in S$ where $s \neq s'$, there exists a course $c \in C$ such that: 
\begin{align}
    c_{s', c} - (1 - c_{s,c}) &\geq \alpha_{s, s'} \\ 
    p_{s,c} &> p_{s', c} \\  
    d_{s,c} &> d_{s', c} \\ 
    \alpha_{s, s'} &\implies \sum_{c\in C} c_{s,c} < \sum_{c\in C} c_{s',c}
\end{align}
Note that each condition of the fairness constraint is written separately, whereas, in Section 6, we test each condition of the fairness to ensure its feasibility. 

We encode new constraints outside of the original problem to adjust to the introduction of binary variables. 
To ensure that the IP algorithm does not create conflicting schedules, we define the following constraints for feasibility. 

First, we ensure that a single assignment is made per course per student (i.e., a student is enrolled in one instructor's lecture section). 
\begin{equation}
    \sum_{c \in E_i} a_{c, i, s} \leq 1 \text{ } \forall i \in I, \forall s \in S \\ 
\end{equation}

Next, all units $u \in U$ should have a corresponding instructor $i$ and room $r$ assigned to this unit. 
Additionally, no student should be assigned to a unit that does not exist. 
If a student is assigned to a unit for course $c$ and instructor $i$, then all corresponding assignments and enrollment values for lecture $(c,i)$ should reflect that as well. 
\begin{align}
    u_{c, i, r, t} &= u_{i, u} \\ 
    u_{c, i, r, t} &= u_{r, u} \\ 
    (1 -  u_{c, i, r, t}) &\leq (1 - u_{s, u}) \\ 
    u_{s, u} &\leq a_{c, i, s} \text{ } \forall s \in S \\ 
    u_{s,u} &\leq c_{s, c} \text{ } \forall s \in S
\end{align}

We also ensure that every lecture $(c, i) \in L $meets once per day on separate days. 
In other words, we make sure that a lecture does not meet only on one day of the week. 
Additionally, we ensure the number of lectures taught in a period does not exceed the total number of available rooms to encourage lectures to occur concurrently. 
\begin{align}
    \sum_{r \in R} \sum_{t \in T_d} u_{c, i, r, t} &\leq 1 \text{ } \forall d \in D \\ 
    \sum_{u\in U} u_{c, i, r, t} &\leq |R| \text{ } \forall t \in T
\end{align}

Feasibility is ensured by making sure each unit has a corresponding assignment and lecture and the relationship between all three decision variables is established.   
\begin{align}
    u_{c, i, r, t} &\leq l_{c, i} \\ 
    a_{c, i, s} &\leq l_{c, i}  
\end{align}
For all lectures $(c, i) \in L$, every assignment $a_{c, i, s} \in A$ should have at least more than one unit. 
\begin{align}
    a_{c, i, s} \implies \sum_{\forall r \in R} \sum_{\forall t \in T} u_{c, i, r, t}  &\geq 1 \text{ } \forall s \in S 
\end{align}
Finally, every student $s \in S$ should be enrolled in a course if and only if an assignment and lecture exist. 
\begin{align}
    c_{s, c} &\leq a_{c, i, s} \\ 
    c_{s, c} &\leq l_{c, i} \\ 
    a_{c, i, s} + l_{c, i} - 1 &\leq c_{s, c} 
\end{align}

We have two objective functions, each used separately for the two problem formulations we introduced in section 3. 
For the HSSP, we define the objective as the maximization of assignments: 
\begin{align}
    \text{max} \sum_{a \in A} a_{c, i, s} 
\end{align}

Recall the objective of the FHSSP is to minimize the number of assignments that cause student envy. 
We can describe the objective function below: 
\begin{align}
    \text{min} \sum_{s, s' \in S} \alpha_{s, s'} \text{ if $s \neq s'$} 
\end{align}

\section{Experiment}
\subsection{Experimental Setting}
In this section, we evaluated the performance of our algorithm by running experiments on a real-life data set provided by a private high school located in the San Francisco Bay Area, California. 
The data set contained course requests by a total of 295 students over 121 courses. 

All experiments were conducted on a laptop with an Apple M2-Max processor and 32 GB of memory. 
Our IP model was implemented using Google OR-Tools' CP-SAT solver \footnote{\url{https://developers.google.com/optimization/cp/cp_solver}} and was written in Python. 

Due to the large problem size, we tested the algorithm on six roughly equal distinct subsets of students to test whether an optimal solution exists. 
We summarized the basic information of each problem setting we tested in Table \ref{tab:setting}. 
Because we were only provided data on the student's course requests, we generated synthetic data for the instructors and rooms. 
We assumed that i) each course has its own room, and ii) each instructor is eligible to teach a fixed number of courses. 
We also assume that iii) there are 6 periods per school day; iv) each course must meet more than once on separate days of the week; and v) students may take at most 6 courses. 

Other input values used in the algorithm were pre-calculated or randomized due to the lack of information. 
The course priority order was generated by randomly assigning the grade for its prerequisite to each student eligible to take that course. 
The student's preference order was generated by utilizing the student's first and second-choice requests from the data set and assigning degree of interest values of 1 and 0.5, respectively. 
The remaining courses not requested by the student were randomly assigned degree of interest values using a normal distribution within a range between 0 to 0.5. 
The maximum number of units an instructor teaches per week is the course frequency per week $\times$ number of courses that the instructor is eligible to teach. 

For each subset of students, the experiments were organized as the following below: 
\begin{enumerate}
    \item We tested the feasibility of our algorithm to solve the HSSP under a given problem setting. 
    \item Once feasibility was determined, we gradually added each condition of the fairness constraint to the existing model and tested whether an optimal outcome was returned. 
    \item We continue the previous step until all conditions of the fairness constraint are added to the model and tested whether an optimal outcome was returned. 
\end{enumerate}

\begin{table}[]
    \centering
    \begin{tabular}{ll}
    \toprule
    Number of students            & 295 \\
    Number of courses             & 121 \\
    Number of instructors         & 72  \\
    Number of rooms               & 106 \\
    Minimum instructor units   & 0   \\
    Maximum instructor units   & 16  \\
    Minimum student courses    & 5   \\
    Maximum student courses    & 6   \\
    Number of periods             & 6   \\
    Maximum instructor courses & 4   \\ 
    \bottomrule
    \end{tabular}
    \caption{Summary of parameters used in IP model. The last row refers to the maximum number of courses an instructor can teach.}
    \label{tab:setting}
\end{table}

\begin{table*}[]
    \centering 
    \begin{tabular}{ccccccccc}
    \toprule
    Problem setting & Number of students & Number of courses & Number of student envy & Solver time &  &  &  &  \\
    1               & 48             & 5          & 0               & 25.2s       &  &  &  &  \\
    2               & 48             & 5          & 0               & 26.2s       &  &  &  &  \\
    3               & 49             & 5          & 0               & 26.9s       &  &  &  &  \\
    4               & 50             & 5          & 0               & 20.2s       &  &  &  &  \\
    5               & 50             & 5          & 0               & 18.4s       &  &  &  &  \\
    6               & 50             & 5          & 0               & 19.0s       &  &  &  &  \\ 
    \bottomrule
    \end{tabular}
    \caption{Results of experiments for each problem setting. The third column refers to the number of courses assigned to each student.} 
    \label{tab:results}
\end{table*}

\subsection{Experimental Results}
Our experiments showed that, across all problem settings, our algorithm consistently produced a feasible solution for the HSSP and an optimal solution for the FHSSP.
Additionally, there was no instance of student envy created by the algorithm for all problem settings. 
In regards to running time, it took approximately 9 minutes to reach a feasible solution, and an additional 3 min to reach an optimal solution. 
This could become an issue as the problem size increases, in which case a computer with parallel processing is necessary to obtain a solution within a reasonable period of time. 

Overall, our algorithm provided an optimal course schedule that effectively minimized the total student envy function while satisfying the constraints imposed on the model, hence solving the FHSSP.
A summary of the experimental result can be found in Table \ref{tab:results}. 

\section{Conclusion}
In conclusion, we successfully developed an integer programming (IP) model that incorporates fairness constraints (such as envy-freeness) into the course scheduling method of American high schools. 
The introduction of fairness constraints into the high school scheduling problem (HSSP) has shown that this issue can be resolved.
By incorporating the concept of fairness from market design into traditional combinatorial optimization problems, the fairness of the high school course scheduling process has been improved. 
However, as the problem size increases, such as the number of students and courses, it becomes more difficult to obtain an optimal solution. 
Therefore, we look forward to improvements to shorten CPU time and reduce the number of variables in our model. 
We believe that this research can be applied not only in educational settings but also in other fields. 
This research shows that it is possible to improve problems caused by luck, disparity, and lack of transparency.
Finally, we stress the necessity to incorporate human emotions (such as envy and unfairness) into mathematical modeling based on real-life social problems and achieve fair matching in the future. 

\section*{Acknowledgements}
This work was supported by the JST Global Science Campus Experts in Science Information program (NII, IPSJ, JCIOI). We would like to extend our deepest gratitude to Menlo School for providing the course request dataset and valuable insights. 

\bibliographystyle{ieeetr}
\bibliography{mybib}

\end{document}